# Integrated Silicon Photonic Multichannel Optical Hybrid for Broadband Parallel Coherent Reception

Tong Lin[1,2†,*], Yan Fan[1†], Jiao Zhang[3], Wenqi Yu[4], Zhengyu Guo[5], Liu Li[6], Zhigang Xin[3], Mingzheng Lei[3], Ziyang Xiong[1], Haoran Wang[1], Hao Deng[1], Min Zhu[3,7], Shihua Chen[3,8], Junpeng Lu[1,*], and Zhenhua Ni[1,3,8]

[1]School of Electronic Science and Engineering, Southeast University, Nanjing, Jiangsu, 210096, China

[2]Key Lab of Modern Optical Technologies of Education Ministry of China, Soochow University, Suzhou 215006, China

[3]Purple Mountain Laboratories, Nanjing, Jiangsu, 211111, China

[4]Institute for Electronics and Information Technology in Tianjin, Tsinghua University, Tianjin, 300467, China

[5]School of Integrated Circuits, Southeast University, Nanjing, Jiangsu, 211189, China

[6]School of Material Science and Engineering, Southeast University, Nanjing, Jiangsu, 210096, China

[7]School of Information Science and Engineering, Southeast University, Nanjing, 210096, China

[8]School of Physics, Southeast University, Nanjing, 21119, China

**ABSTRACT:** We design and demonstrate a monolithically integrated silicon photonic multichannel optical hybrid for versatile broadband coherent reception, addressing the critical limitations of current wavelength multiplexed systems in scalability and power efficiency. The device combines a phase-compensated 90° optical hybrid with four robust three-stage Mach-Zehnder interferometer lattice filters, enabling 34-port functionality (two inputs and 32 outputs) for simultaneous analog and digital signal processing. Leveraging multimode interferometer designs, the chip achieves a broadband response with sub-dB passband uniformity across eight 200 GHz-spaced wavelength channels, while maintaining phase errors below 4° over a 13.5 nm bandwidth (1539–1552.5 nm) with only 2.5 mW thermal tuning power. Experimentally, we validate its parallel-processing capability through RF channelizer reception (showing an average spurious-free dynamic range of 80.8 dB·Hz$^{2/3}$ and image rejection ratio of 33.26 dB) and coherent optical communication (achieving 1.024 Tb/s data rate for 32-QAM signals with bit error rates far below the 20% SD-FEC threshold). The scheme enhances system performance with fully passive wavelength multiplexing integration, supporting high-fidelity uniformity and projecting scalability to 1.468 Tb/s. This work promises advancements in high-performance optoelectronic devices for next-generation AI-driven data centers and 5G-XG networks.

## 1. INTRODUCTION

The rapid advancement of generative artificial intelligence (AI) [1-3], 5G-XG networks, autonomous driving, and the Internet of Things [4] has triggered explosive growth in multi-modal data streams. These data floods permeate hyperscale data centers, unmanned vehicles, and mobile platforms, propelling global data traffic at a Compound Annual Growth Rate of ~60% [5, 6]. To address this exponential capacity demand, photonic integrated circuits (PICs) have emerged as the key enabler. By leveraging multidimensional multiplexing photonic systems-on-chip integrated with coherent detection technologies, PICs fully exploit their inherent advantages of ultrabroadband bandwidth, high spectral efficiency [7], high sensitivity, and scalability for mass deployment.

Among diverse material platforms for PICs [8, 9], silicon photonics [10] have become a pivotal choice due to its exceptional compatibility with standard complementary metal-oxide-semiconductor (CMOS) processes and its capability for high integration density. Within silicon photonic systems, wavelength-division multiplexing (WDM) [11] stands as the primary multiplexing technology, providing an essential solution to meet the escalating bandwidth demands in coherent optical communications. A typical silicon photonic WDM transceiver operates as follows: on the transmitter side, multiple wavelength carriers are individually encoded using complex modulation formats [12] [STQEreviwer] such as single-sideband (SSB) modulation, quadrature phase-shift keying (QPSK), or quadrature amplitude modulation (QAM). These modulated signals are then combined through a multiplexer (Mux) for output transmission via fiber [13] or free space. On the receiver side, a demultiplexer (DeMux) separates the incoming composite signal into its constituent wavelength channels. Each channel is subsequently delivered to a 90° optical hybrid (OH), a quadrature demodulation stage for data retrieval [14]. Both the WDM Mux/DeMux and the 90° OH are cornerstone passive components within coherent detection systems. They enable the critical extraction of both amplitude and phase information from the signal, supporting high single channel data rates of 100 Gb/s and beyond [9, 15, 16].

Current silicon photonic coherent detection systems predominantly rely on hybrid integration of optical hybrids (OHs) alongside discrete WDM devices, such as Z-Block components and silica-based PLC arrayed waveguide gratings. Adopting monolithic integration instead would substantially reduce the overall form factor while simultaneously minimizing packaging parasitics, system complexity, and cost. However, current chip-scale implementations face critical limitations. State-of-the-art silicon photonic WDM coherent receivers are constrained to 4-5 wavelengths [17-19], significantly underperforming compared to discrete photonic counterparts (i.e., 25 wavelength channels in [20], 34 wavelength channels in [10]. This performance gap arises from two challenges. One is the large channel non-uniformity and excessive tuning power caused by phase-sensitive arrayed waveguide gratings [17] or multiple microring resonators (MRRs) [18] inside. The other is the sophisticated quasi-WDM strategy relying on equal power splitting among multiple different local oscillators (LOs) and electrical filtering [19], further reducing received signal powers.

In this study, we propose and experimentally demonstrate a silicon photonic multichannel wavelength demultiplexed optical hybrid (WDM-OH) and show its multifunctional capability of parallel-processing wideband analog and digital signals. By leveraging robust multimode interferometer (MMI) designs, the constructed 90°OH integrated with four WDM filter array exhibits high phase and amplitude consistency. The 90° OH maintains less than 4° phase error from 1539 nm to 1552.5 nm, requiring merely 2.5 mW thermal tuning power. The Mach-Zehnder interferometer (MZI) lattice filters enable a fully passive 200 GHz-spaced WDM filter array spanning eight channels with sub-dB passband uniformity, eliminating the need for active tuning in conventional phase-sensitive arrayed waveguide gratings or MRRs. These enable wideband RF channelizer reception and optical fiber communication, achieving a 1.024 Tb/s 25 GBaud 32-QAM coherent reception in experiments. This work advances large-scale WDM coherent receiver systems on a fully monolithic platform, offering high integration density, minimal packaging parasitics, and reduced system complexity by leveraging reliable CMOS-compatible fabrication. These advancements contribute to next-generation optical communication and interconnects systems in generative AI, 5G-XG networks, ultra-wideband radars, and the Internet of Things.

## 2. RESULTS AND DISCUSSION

We propose a monolithically integrated 34-port silicon photonic WDM-OH chip combining a phase-compensated 90°OH with four robust three-stage MZI lattice filters, achieving high-performance wavelength demultiplexed quadrature optical mixing. Figure 1(a) displays a schematic illustration of the chip with two input ports (i.e., signal, and local oscillator) and 32 output ports (4 identical sets of 8-wavelength channels). As shown in Fig. 1(b), the 90° OH resembles a traditional Chinese knot, comprising three 2×2 MMIs, one 1×2 MMI, and four sets of bent waveguides. By utilizing intrinsic $\pi/2$ output phase profile of 2×2 MMIs and 1×2 MMI's symmetry configuration [21], the Chinese knot-type 90° OH maintains relative phase shift among four outputs in a broad bandwidth. TiW electrodes are deposited on bent waveguides to compensate for phase errors caused by fabrication defects, while large-radius bends (each bent waveguide is made of three $\pi/4$ circular bends with a 70 µm bending radius) are adopted to suppress thermal crosstalk. To mitigate chromatic dispersion, we implement MMIs instead of conventional directional couplers, significantly reducing wavelength-dependent dispersion. Figure 1(c) shows insertion losses and imbalances of these two MMIs with a bandwidth of 100 nm: the variations are less than 0.3 dB (0.5 dB) for the C-band (100nm bandwidth). The four outputs (I$^+$, I$^-$, Q$^+$, Q$^-$) of the 90° OH are meticulously aligned to the input ports of the WDM for wavelength demultiplexing. Each WDM consists of the three-stage MZI lattice filter made of 2×2 MMIs and connected with decremental S-bends to de-interleave adjacent wavelength carriers. Notably, the 32 optical paths from the 90° OH to the output TE-polarized grating couplers are matched to minimize temporal delays before entering into photodetectors.

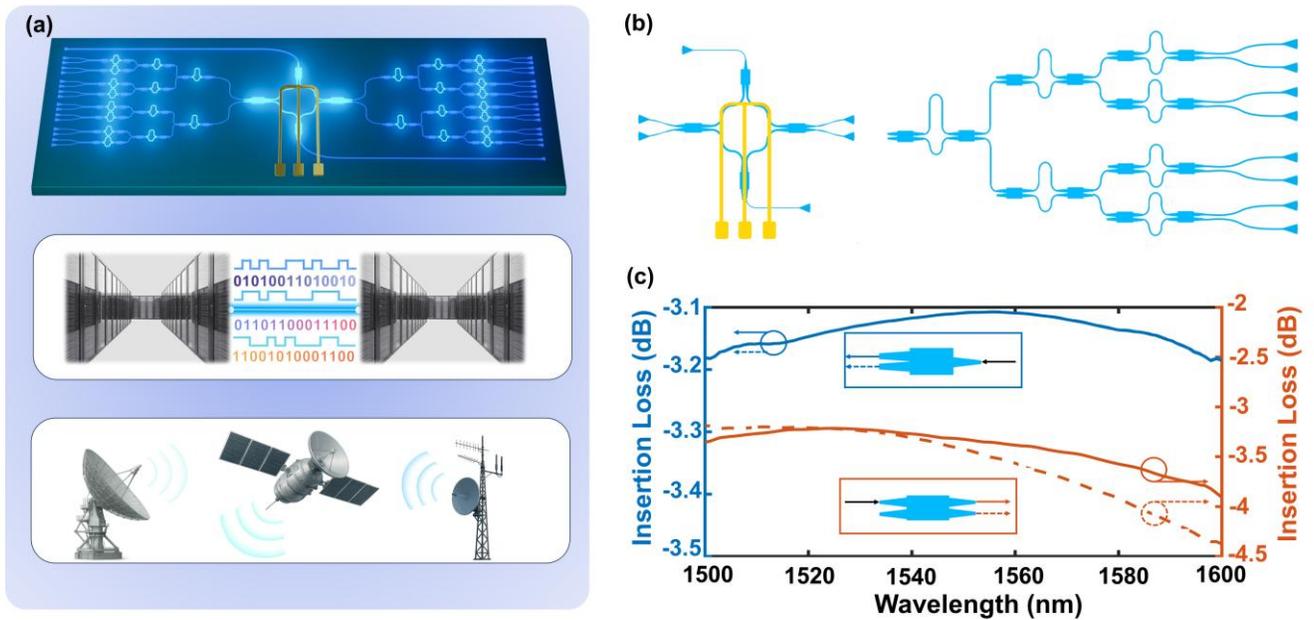

**Figure 1.** (a) Artist 3D schematic of the 34-port silicon photonic WDM-OH with analog and digital signal processing applications. (b) Schematic diagram of the phase-compensated 90° OH component and one robust 1×8 MZI-lattice WDM filter with high channel uniformity. (c) The simulated insertion losses of 1×2 MMI and 2×2 MMI for the wavelength from 1500 nm to 1600 nm.

We demonstrate the excellent phase and amplitude consistency of the fabricated chip-scale WDM-OH. The chip utilizes a 220-nm-thick silicon layer cladded with oxide layers on an SOI wafer, fabricated from a Multi-Project Wafer run provided by Applied Nanotools Inc. Figure 2(a) shows the packaged device with single mode fiber arrays for optical interface and wire bonds for electrical connections. The device's entire footprint is about 18.6 mm² (primarily for waveguide routing). Phase characterization employs a time-domain analysis method, eliminating the need for extra unbalanced interferometric structures [22]. In the experimental setup [Fig. 2(a)], dual-tone signals (100 MHz frequency-detuned from a 1550 nm optical carrier) generated by an acousto-optic modulator are injected into the SIG and LO ports. The resulting in-phase (I) and quadrature (Q) sinusoidal waveforms are captured by a mixed-signal oscilloscope (MSO), enabling extraction of I-Q phase errors from time-domain waveform delays. By changing the optical carrier wavelength, the phase errors of all the output ports are mapped subsequently. To minimize fabrication-induced phase deviations, a 0.85 V DC bias is applied to one bend of the 90° OH, consuming only 2.5 mW static power. Sequential wavelength scanning results [Fig. 2(b)] confirm that the phase errors between $I^+$ and $Q^+$ outputs for the 16 WDM channels remain below 4° over a bandwidth of 12 nm, compliant with the optical internetworking forum standards [22].

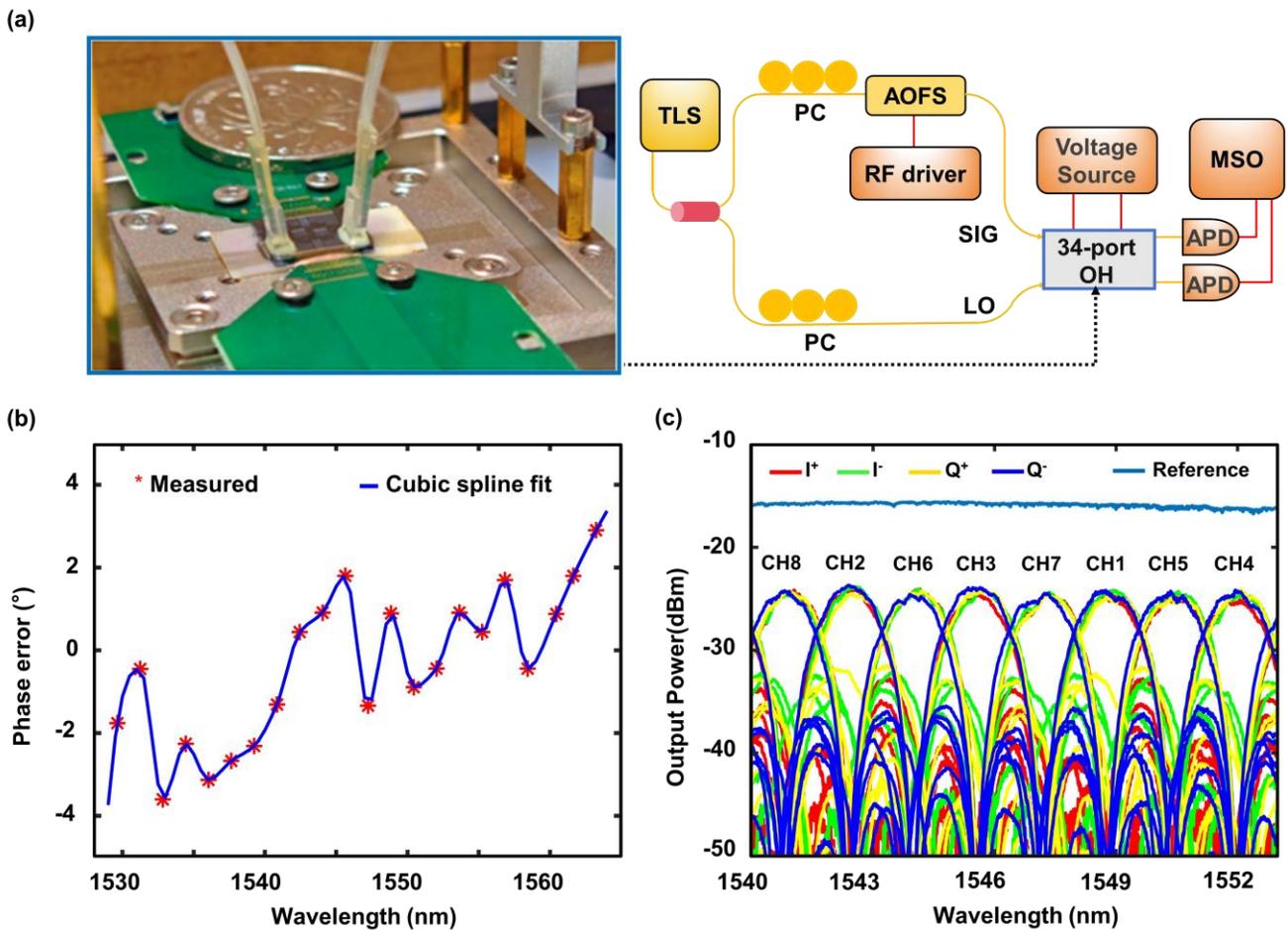

**Figure 2.** (a) An optical microscope image of the packaged 34-port WDM-OH chip with wire bonds and fiber arrays (the left panel). The right panel shows the experimental setup for phase error characterizations (PC:

Polarization Controller, AOFS: acoustic-optical frequency shifter, MSO: RIGHO MSO8000A, APD: avalanche photodetector). (b) Extracted phase errors between the $I^+$ and $Q^+$ branches of the chip-scale WDM-OH for the wavelength spanning from 1540.5 nm to 1552.5 nm. The red asterisks denote the 8 channel peak wavelengths' phase error between $8I^+$ and $8Q^+$ channels. (c) Measured transmission spectra for 32 output ports of the chip-scale WDM-OH and the reference waveguide. The (non-) adjacent channel crosstalk is (14.2) 9.3 dB.

The four DWDM filters are highly tolerant to fabrication imperfections and work in an all-passive manner without post trimming. As depicted in Fig. 1(b), each filter utilizes a three-stage cascaded MZI lattice configuration. To realize a WDM channel spacing of 1.6 nm, the FSRs of successive MZI stages are designed to double progressively ($\Delta\lambda_{1/2/3} \approx 3.2/6.4/12.8$ nm), which corresponds to the offset length differences ($\Delta L_{1/2/3} \approx 178.948/89.474/44.737$ µm). With extra offset length differences, the Gaussian-like passbands [23] are generated at each wavelength channel. Figures 2(c) presents the optical transmission spectra of all 32 output ports and the U-shaped reference waveguide. Clearly, each channel exhibits a 3-dB bandwidth of 120 GHz. After compensating for the intrinsic 6 dB loss from the optical hybrid (OH), the insertion losses across all 32 channels are summarized in Table 1, ranging from –1.37 dB (minimum) to –2.44 dB (maximum), yielding inter-channel uniformity of approximately 1.07 dB. Furthermore, the imbalances of the same path (ISP) are less than 0.96 dB. This high uniformity significantly relaxes the requirements for IQ equalization algorithms in coherent receivers. Critically, the high uniformity of our WDM design eliminates the need for thermal tuning to maintain wavelength alignment, thereby minimizing power consumption and reducing design complexity compared to conventional microring-based alternatives.

Table 1. Insertion losses of the chip-scale WDM-OH's 32 channels.

| Path | Insertion loss | | | | | | | | |
|------|-----|-----|-----|-----|-----|-----|-----|-----|-----|
|      | CH1 | CH2 | CH3 | CH4 | CH5 | CH6 | CH7 | CH8 | ISP |
| $I^+$ | -2.18 | -1.98 | -2.26 | -2.21 | -2.11 | -1.86 | -2.08 | -1.92 | 0.4 |
| $I^-$ | -1.95 | -1.79 | -2.07 | -1.68 | -1.78 | -1.89 | -2.01 | -2.31 | 0.53 |
| $Q^+$ | -2.33 | -2.31 | -2.08 | -1.37 | -2.11 | -2.11 | -2.07 | -2.08 | 0.96 |
| $Q^-$ | -1.89 | -1.58 | -1.83 | -1.53 | -1.63 | -2.44 | -2.20 | -2.02 | 0.86 |
| **ISW** | 0.44 | 0.73 | 0.43 | 0.84 | 0.48 | 0.58 | 0.12 | 0.42 | |

ISW: Imbalance of the same wavelength

ISP: Imbalance of the same path

To validate the broadband analog signal parallel processing capacity of our chip, we demonstrate the coherent RF channelized reception [20] using the experimental setup in Fig. 3(a). A wavelength-tunable optical carrier is split by a 3 dB fused fiber coupler: one path serves as the LO; the other is single-sideband frequency shifted by an IQ modulator loaded with RF signals from an arbitrary waveform generator (AWG). This AWG enables flexible selection of $\pm 1^{st}$-order sidebands by adjusting the I/Q phase difference [24, 25]. We first characterize the reception channel linearity by measuring the spurious-free dynamic range (SFDR). Dual-tone RF signals at 0.5 GHz and 0.6 GHz (-1$^{st}$ sidebands) serve as the signals under test (SUTs), generating third-order intermodulation distortion (IMD3) components at 0.4 GHz and 0.7 GHz. As shown in Fig. 3(b), the power slopes of fundamental signals versus IMD3 components yield an SFDR of 81.2 dB·Hz²/³ (background noise: -120.53 dBm/Hz). For image rejection ratio (IMRR) characterization, a 0.8 GHz tone (-1$^{st}$ sideband) is applied as the SUT, while the +1$^{st}$ sideband acts as the image frequency. The SUT and LO are fed into the WDM-OH for demodulation, followed by photodetection to obtain in-phase/quadrature intermediate frequency (IF) signals at 0.8 GHz. These signals are processed by an electrical hybrid (EH) for image rejection [26], achieving >30 dB suppression when switching between sidebands. The real-time spectrum in Fig. 3(c) confirms an IMRR of 31.62 dB in Channel 6. By sweeping optical wavelengths across all 8 channels (Fig. 3(d)), we obtain an average SFDR of 80.8 dB·Hz²/³ and average IMRR of 33.26 dB, demonstrating high channel uniformity and crosstalk suppression performance.

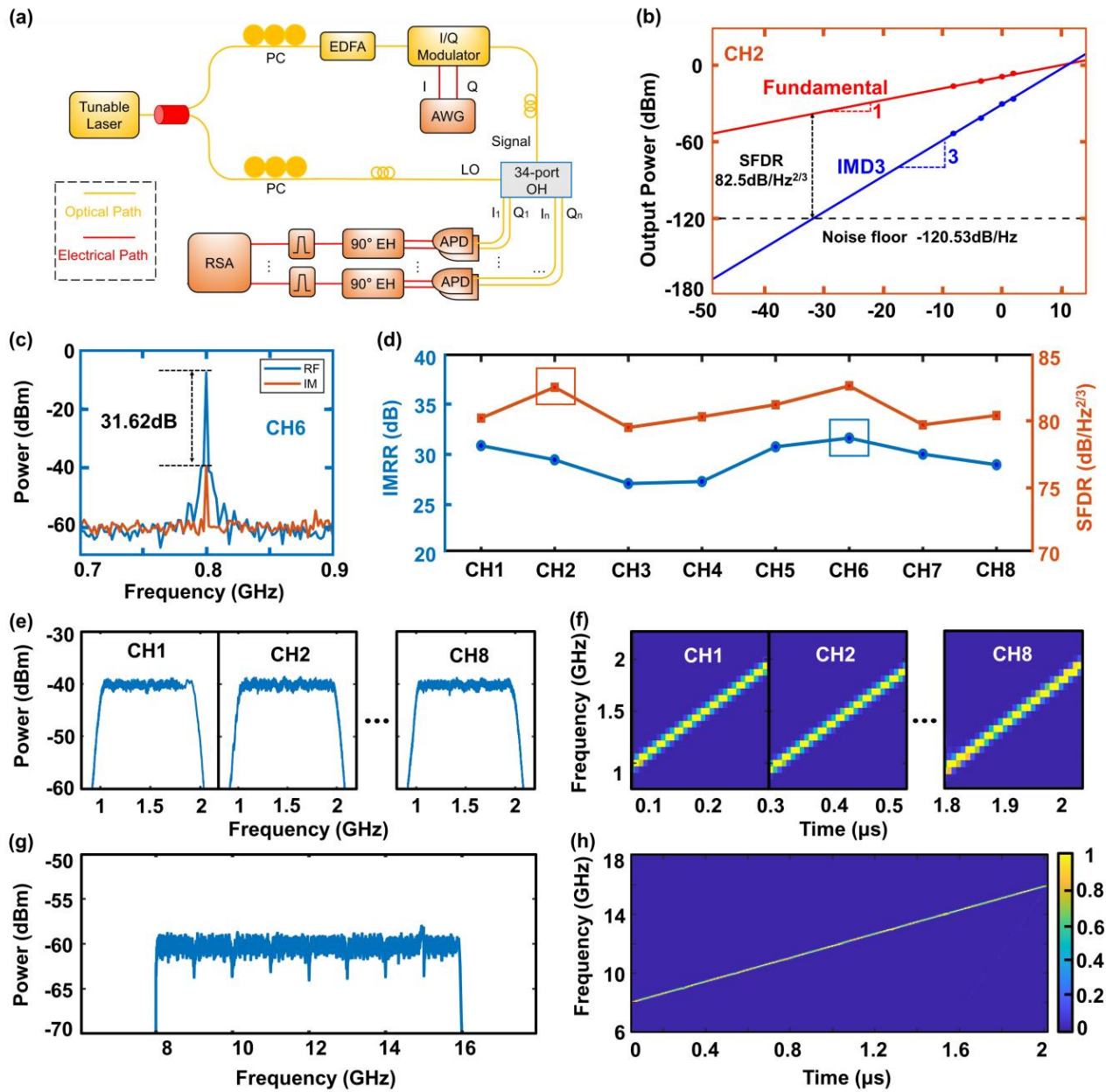

**Figure 3.** (a) The schematic diagram of a RF channelizer coherent receiver system using the fabricated 34-port wavelength multiplexed 90° OH chip (EDFA: Erbium-Doped Fiber Amplifier, AWG: Keysight 8195A, Electrical Hybrid (90° EH), electrical bandpass filter, RSA: real-time spectrum analyzer, RIGOL RSA3045N). (b) Output powers of the fundamental and IMD3 components of Channel 2 versus the different input RF powers measured at a resolution bandwidth (RBW) of 100 kHz. (c) Channelized output spectra of the SUTs at Channel 6. (d) The measured IMRRs and SFDRs for all 8 channels. LFM channelized at each channel (after a 3rd-order Bessel electrical bandpass filter (Freq.:1-2 GHz; 3dB BW: 1 GHz)) (e) in the RF frequency domain; (f) in the time domain. (g) The reconstructed output spectra of the wideband SUT. (h) The frequency-to-time relation of the reconstructed LFM SUT.

We further evaluate the multi-band RF signal reception capacity utilizing the proposed 34-port WDM OH chip and provide a simulation-based demonstration. This multi-band linear frequency modulated (LFM) RF signal, generated in Optisystem software, operates at a carrier frequency of 12 GHz with a bandwidth spanning 8–16 GHz, a pulse period of 2 μs, and a duty cycle of 0.9. This signal is loaded onto a signal optical frequency comb (OFC) with a FSR of 200 GHz through a SSB IQ modulator. Simultaneously, a LO OFC (201 GHz FSR) with a carrier frequency offset ($f_{0,loc}=f_{0,sig}+7$ GHz) is injected into the channelization receiver. The original 8 GHz bandwidth LFM signal is channelized into 8 channels with a 1 GHz narrowband, in which the down-converted signals are processed consistent with the method described previously and fall within the same IF range across all channels. Figures 3(e) and 3(f) respectively depict the spectrum and time-frequency diagrams of the channelized signals, confirming uniform spectral slicing over 8 channels. After the final up-conversion process, the reconstructed signal exhibits spectral characteristics nearly identical to the original LFM signal as shown in Fig. 3(g). Crucially, the time-frequency diagram (Fig. 3(h)) shows single-frequency dominance at each time slice, demonstrating effective suppression of image frequencies by the image-reject mixer. This validates the channelization scheme's capability for high-fidelity crosstalk suppression of wideband LFM signals.

We conduct a coherent communication system experiment to validate the device's capability of parallel-processing broadband digital signals. Figure 4(a) illustrates the experimental setup for the back-to-back coherent transmission and reception using the WDM-OH chip. At the transmitter side, an AWG operating at 92 GSa/s generates signals (offline processing details are given in the Methods section) with diverse modulation formats (e.g., QPSK, QAM), which are encoded onto an optical carrier via a 35 GHz electro-optic modulator; an erbium-doped fiber amplifier (EDFA) subsequently amplifies the modulated signal prior to chip coupling. A frequency-offset laser serves as the LO. At the receiver side, the silicon photonic chip simultaneously performs optical mixing of the signal/LO and channel demultiplexing. Balanced photodetectors convert the demultiplexed optical signals to electrical domain, followed by amplification and digitization using an 80 GSa/s DSO. Offline signal processing employs custom-developed digital signal processing (DSP, its block diagram is depicted in our previous work [27]). The Gram-Schmidt orthogonalization procedure is employed to compensate for I/Q imbalance. Subsequently, a blind equalizer based on the constant modulus algorithm with 37 taps is utilized to mitigate linear polarization crosstalk. Carrier phase recovery is performed through a two-stage process comprising principal component-based phase estimation and blind phase search. In addition, a data-aided MIMO structure Volterra nonlinear equalizer [28] is introduced to suppress nonlinear impairments introduced by the fiber transmission, photoelectric detection and amplification links while this equalizer is updated via least mean squares adaptation. Finally, bit error rates (BERs) of the recovered signals are calculated to quantify system performance.

Figure 4(b) displays the BER heat map for multiple SUTs across 16 wavelength channels (8 in-phase $I^+$ and 8 quadrature $Q^+$ channels). These include 25 GBaud QPSK, 8-QAM, 16-QAM, and 32 GBaud 16-QAM formats. For the 25 GBaud QPSK signal, all 8 channels achieve error-free operation (BER = 0). The 25 GBaud 8-QAM and 16-QAM signals exhibit BERs ranging from $9.1\times10^{-4}$ to $3.1\times10^{-3}$, and $1.31\times10^{-3}$ to $4.07\times10^{-3}$ respectively, with 7 channels below the HD-FEC threshold of $3.8\times10^{-3}$. All channels for the 32 GBaud 16-QAM signal show BERs < $6.53\times10^{-3}$, significantly lower than the 20% SD-FEC threshold of $2\times10^{-2}$ [27, 29]. Consequently, coherent reception at a total data rate of 1.024 Tb/s (32 GBaud × 4 bits/symbol × 8 wavelengths) is experimentally validated. Figure 4(c) further demonstrates demodulated constellation diagrams for 25/32 GBaud 16-QAM signals, with all channels exhibiting error vector magnitudes (EVM) below 22.5%. Notably, the current aggregate data rate is constrained solely by the sampling rate limitation of the available DSO. To assess the fundamental capacity limit, we simulate projected BERs for higher data rates under identical signal-to-noise ratio (SNR) conditions. Least-squares fitting results for Channel 1 (Fig. 4(d)) indicate performance saturation near 51.93 GBaud due to cumulative nonlinear effects. This projects a maximum achievable total data rate of 1.468 Tb/s across all channels, as quantified in Fig. 4(e).

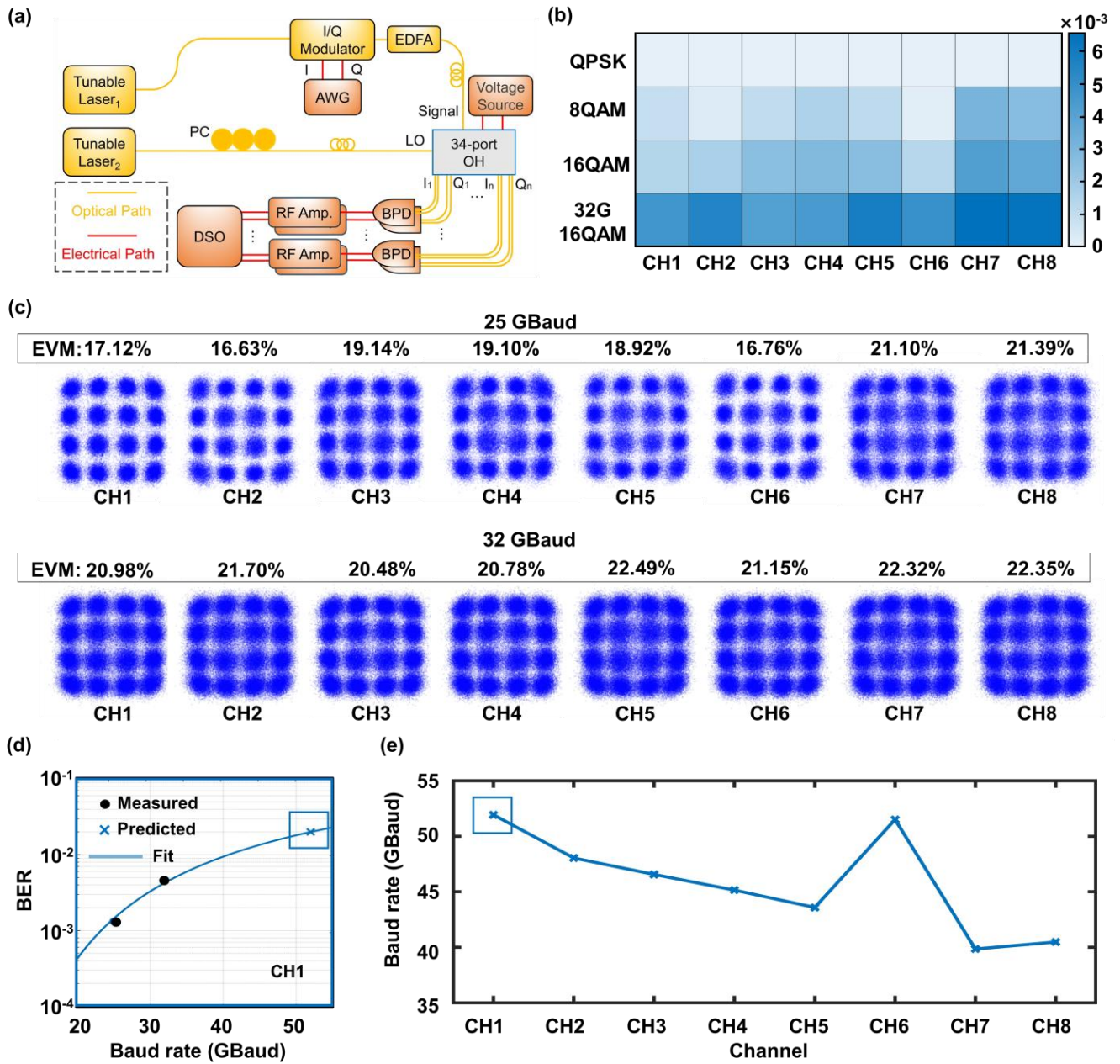

**Figure 4.** (a) Coherent communication system experimental diagram using the fabricated 34-port wavelength multiplexed 90° OH chip ( DSO：Keysight DSAV334A) ; yellow and red lines represent optical and electrical paths respectively. (b) Measured BER heat map of SUTs (25 GBaud QPSK, 8-QAM, 16-QAM, and 32 GBaud 16-QAM) for 8 channels. (c) The reconstructed constellations of 25 GBaud 16-QAM, and 32 GBaud 16-QAM signals for 8 channels under a 20 dB OSNR and 2 dBm output optical power right after the chip. (d) The fitted and projected BERs of the CH1 signal from 20 to 55 GBaud rates. (e) The maximum achievable GBaud rates for 8 channels under the same SNR.

We summarize the chip performance of various multichannel WDM receivers based on SOI or other material platforms in Table 2, which can be categorized into two types. For intensity modulation direct detection (IMDD) schemes, the highest data rate tolerating the SD-FEC limit ($2\times10^{-2}$) reaches 1.29 Tb/s (8 channels, O-band) [30], where interleaved angled-MMIs (AMMIs) function as the DeMux. However, AMMIs exhibit high sensitivity to fabrication imperfections, evidenced by a 5.6-dB inter-channel uniformity. This limitation arises from irregular angles exacerbating mask discretization effects, rendering AMMIs unsuitable for standard mask-writing strategies [31, 32]. To circumvent phase sensitivity, arrayed waveguide gratings are built on low index contrast material platforms (i.e., Polymer [33], $Si_3N_4$ [34]). Yet, this approach inevitably entails significantly larger footprints. While MRR [35] and Microdisk [36] arrays can be very compact, they introduce new trade-offs: dependence on real-time photodetection feedback and digital control circuitry increases system complexity. Notably, such active tuning demands substantial power—about 5.5 mW per channel [36]. Emerging solutions like phase-change materials [37] offer energy-efficient wavelength trimming. Their reliance on one-to-one calibration remains prohibitively time-consuming for scalable deployment. Nevertheless, the IMDD scheme alone exclusively extracts amplitude information. In contrast, coherent receivers not only recover both amplitude and phase information of optical carriers but also achieve much higher data rates than IMDD schemes. For instance, an InP coherent optical engine [38] provides 800 Gb/s net data rate per channel using free-space coupling optics, though this approach introduces additional packaging complexity and cost. Integrated WDM receivers show inferior performances [39] due to rising crosstalk and channel uniformity of integrated demultiplexers. To mitigate crosstalk, a coherent receiver using the quasi-WDM scheme [19] utilizes a dual-stage filtering strategy- utilizing both photodetector frequency roll-off and electrical low-pass filtering. They show a nominal aggregate net data rate of 1.22 Tb/s (10 physical channels) with the BER below 7% hard-decision FEC threshold of $3.8 \times 10^{-3}$. Nevertheless, this scheme necessitates equal power splitting of signals across multiple wavelength-shifted LOs, fundamentally constraining its scalability due to link power budget limitations. It is highlighted that our device shows the highest data rate per wavelength channel, which is only limited by the available instruments. The projected aggerate data rate can be up to 1.468 Tb/s given the same SNR condition.

Table 2. Experimental performance of representative WDM receivers

| Ref. | Receive type | Number of channels | DeMux | Bit rate [Tbps] | Size [mm$^2$] | Platform |
|------|--------------|--------------------|-------|-----------------|---------------|----------|
| [33] | IMDD | 8λ | Arrayed waveguide grating | 0.8 | >19.5×6.25 | InP-Polymer |
| [30] | IMDD | 8λ | A-MMI +MZI | 1.29 | 5×5 | SOI |
| [35] | IMDD | 4λ | Dual-MRRs | 0.2 | 14.04 | SOI |
| [36] | IMDD | 32λ | Microdisk | 0.576 | 5.5×7.5 | SOI |
| [34] | IMDD | 40λ | Arrayed waveguide grating | 0.04 (one channel) | 6.5×5 | Si-Si$_3$N$_4$ |
| [37] | IMDD | 4λ | MRR | 0.4 | / | SOI-SbSe |
| [39] | Coherent | 8λ | Arrayed waveguide grating | 0.2 | >25×12.5 | InP-Silica |
| [38] | Coherent | 2λ×2 Pol. | Free-space optics | 1.6 | >42×42 | InP |
| [19] | Coherent | 5λ×2 Pol. | Quasi-WDM | 1.12 | 10.8 | SOI |
| This work | IMDD/ Coherent | 8λ | MZI lattice filter | 1.024 | 7.24×2.57 | SOI |

## 3. CONCLUSIONS

In summary, we proposed and experimentally validated a high-performance silicon photonic multichannel WDM-OH, setting a new benchmark in monolithic integration for versatile broadband coherent reception. The 34-port chip integrates a phase-compensated 90° OH with four MZI-based WDM filters, achieving sub-dB channel uniformity and low phase error (<4°) over an 8-channel 200 GHz grid, with minimal thermal tuning power (2.5 mW). Key system-level demonstrations include RF channelizer reception with an average SFDR of 80.8 dB·Hz²ᐟ³ and IMRR of 33.26 dB; Coherent digital communication supporting back-to-back data capacity of 1.024 Tb/s 32-QAM signals across 8 wavelengths, with BERs < 6.53×10⁻³. Monolithic architecture eliminates the need for discrete WDM components, reducing form factors, packaging parasitics, and system complexity while outperforming state-of-the-art silicon WDM receivers constrained to 5 wavelengths. This work paves the way for scalable, energy-efficient coherent systems in ultra-wideband radars and AI hyperscale data centers with future directions including inverse design [40, 41] for compactness and monolithic integration with Ge photodiodes [42] to further enhance bandwidth and sensitivity.

## METHODS

**Offline DSP initialization:** At the transmitter side, a pseudo-random binary sequence (PRBS) of length 262,144 ($2^{18}$) is generated and mapped to M-QAM symbols. The sequence undergoes two-fold upsampling followed by pulse shaping using a root-raised-cosine filter. A synchronization header is appended to facilitate signal synchronization at the receiver. The resulting waveform is synthesized by an arbitrary waveform generator with a 92 GSa/s sampling rate, producing the in-phase and quadrature components of the electrical baseband signal.

**BER fitting model:** Due to the rate limitation of the AWG, the current BER remains significantly below the threshold. To perform predictions, we first map the BER for each channel at different symbol rates to the corresponding Q-factor and adopt $Q^2$ as the fitting target, establishing a channel degradation model. The Q-factor for each channel is assumed to vary with the symbol rate $R$ according to:

$$Q^2(R) = \alpha_c + \frac{K}{R} \tag{1}$$

where $K$ is a system-level parameter determined by the measured BER across all channels, describing Q-factor degradation as the symbol rate increases. By minimizing the residual sum, we calculate $K \approx 225$.

The channel-specific parameter $\alpha_c$ is derived using Q-values measured at two symbol rate points, expressed as:

$$\alpha_c = mean_i \left( Q_i^2 - \frac{K}{R_i} \right) \tag{2}$$

Given the BER threshold of $2\times10^{-2}$, we determine the target $Q_{tgt}^2$. Substituting this into Equation (1) allows us to solve inversely for the predicted symbol rate ($R_{pred.}$):

$$R_{pred.} = \frac{K}{Q_{tgt}^2 - \alpha_c} \tag{3}$$

**ASSOCIATED CONTENT**

**AUTHOR INFORMATION**

**Corresponding Authors**

Tong Lin – School of Electronic Science and Engineering, Southeast University, Nanjing, China

E-mail: lintong@seu.edu.cn

Junpeng Lu –School of Electronic Science and Engineering, Southeast University, Nanjing, China

E-mail: phyljp@seu.edu.cn

**Author Contributions**

T.L., W. Y., and Y.F. conceived the idea and designed the experiments. Y.F., Z. G., Z.X., H.W., and H. D. simulated the devices. Y.F. and T.L. conducted device characterizations. F.Y., J.Z., Z. X., and M.L. performed system demonstrations. Y.F., T.L., and L. L.  analyzed the experimental data. T.L. and F.Y. wrote this paper. All authors contributed to the discussion and supervision of the paper.

**ACKNOWLEDGMENTS**

This work is supported by National Natural Science Foundation of China (62105061, 12374301, 62225404), Jiangsu Provincial Frontier Technology Research and Development Program (BF2024070), the National Key Research and Development Program of China (2024YFA1210500), and the Key Lab of Modern Optical Technologies of Education, Ministry of China, Soochow University. The authors thank SJTU-Pinghu Institute of Intelligent Optoelectronics for their assistance with the chip packaging.